\documentstyle[twocolumn,aps]{revtex}

\newcommand{\ie}{{\it i.e.}}

\begin{document}

\twocolumn[\hsize\textwidth\columnwidth\hsize\csname %
@twocolumnfalse\endcsname

\draft
\title{Theory of Josephson tunneling along the c-axis of YBCO}
\author{Robert Haslinger and Robert Joynt}
\address{Department of Physics and Applied Superconductivity Center\\
University of Wisconsin-Madison \\
1150 University Avenue \\
Madison, WI 53706 \\}
\date{\today}
\maketitle

\begin{abstract}

The existence of Josephson tunneling has been 
demonstrated between YBa$_2$Cu$_3$O$_{7-\delta}$ and Pb with the 
current flowing along the c-axis of YBa$_2$Cu$_3$O$_{7-\delta}$.  
This is presumed to come from an $s$-wave component
of the superconductivity in YBa$_2$Cu$_3$O$_{7-\delta}$, perhaps induced
by the orthorhombic distortion.  This hypothesis
by itself appears to be in contradiction 
with experiments on twinned samples whose tunneling
current does not follow random statistics.  We present 
a theory which depends on a competition between
the intertwin $d$-wave coupling and a relatively enhanced $s$-wave 
component on the surface.  This theory appears to explain, at least 
qualitatively, all observations made to date.

\end{abstract}

\vspace*{\baselineskip}

]

Understanding the nature of the order parameter
is one of the main challenges in the theory of high-T$_c$
superconductivity.  In YBa$_2$Cu$_3$O$_{7-\delta}$ (YBCO),
Josephson tunneling experiments with current flowing mainly
in the a-b plane \cite{vh} have made it clear that the dominant
component is $d$-wave.  However, a substantial body of 
work has also demonstrated the
existence of Josephson tunneling along the c-axis from 
YBCO to a Pb electrode \cite{dynes}.  This strongly suggests the
presence of an $s$-wave component of the superconductivity
of YBCO.  This is to be expected:
YBCO is orthorhombic and the crystal structure mixes
the $d$-wave and $s$-wave components \cite{lkj}.
This mixing has been adduced in several contexts: it can produce fractional
vortices at grain boundaries \cite{bailey}, and appears to be necessary to
explain features of the microwave response \cite{sridhar}.  
Nevertheless, difficulties arise 
when the tunneling results are considered in detail.  The main problem is
that there is tunneling current even when the sample is twinned.
Twins should occur in equal numbers of $d+s$ and $d-s$, where the
$d$-component is defined with respect to axes fixed in space.  The
net current would then be proportional to to $\sqrt{N_T}$, where
$N_T$ is the total number of twins.  It is actually much larger,
though twinning reduces the net current \cite{oon}.

The theories of c-axis tunneling which have been proposed 
to date have serious drawbacks.  The most detailed is that offered by
Sigrist {\it et al.} \cite{sigrist}.  
Their picture involves no net tunneling from
the twins themselves.  The currents actually come from the twin 
boundaries where the order parameter is complex.  Thus the state
of the sample as a whole actually breaks time-reversal symmetry.
These ideas do not seem to be consistent with the fact that
the current in magnetic field is largely unaffected when the field 
is reversed.  Furthermore, the current should increase when the number
of twins increases.  This is opposite to what is observed.  
In addition, microwave experiments on the junctions indicate that the 
tunneling to Pb is first-order Josephson tunneling.

Some other possibilities suggest themselves.  For example,
a proximity-effect $d$-wave gap may be induced in the Pb
by the high-T$_c$.  However, this is ruled out by the observation that
the supercurrent follows a temperature dependence which
mimics that of the Pb gap.  Also, the coupling to the Pb 
electrode changes sign across a single twin \cite{kouz}.  

The texturing of the 
twins, \ie, the fact that the 
boundaries run predominantly along one direction in space,
might also be adduced as a factor that could produce a net current.
The twin boundaries run along the diagonal of the a-b plane.
Such a texture is equivalent to a rhombohedral distortion, which transforms
according to the B$_2$ representation of the C$_{4v}$ group.  Can such a 
distortion couple $s$-wave (A$_1$) and $d$-wave (B$_1$) order parameters ?
Any such coupling term belongs to the A$_1$ $\times$ B$_1$ 
$\times$ B$_2$ representation.  But A$_1$ $\times$ B$_1$ 
$\times$ B$_2$ = A$_2$ which does not contain the identity reprsentation.
Therefore the introduction of the texture cannot produce a net Josephson 
current.  

Our model takes its cue from the observation that
the magnitude of $I_c R_n$ in general suggests a coupling 
to a high-T$_c$ gap of $\leq 1 meV$, not $20 meV$.        
This suggests that the magnitude of the gap at an (001)
surface is much reduced.  This is consistent with the fact
that, in spite of much effort, photoemission experiments
(with resolutions of order 10 meV) have also never succeeded 
in seeing a gap at this surface.  Furthermore, if this 
reduction is due to disorder, such as surface 
scattering, one would expect that the $d$-wave 
component is relatively much more
suppressed than the $s$-wave \cite{note}.  Similar results
might occur due to an oxygen vacancy concentration gradient \cite{flatte}.
The suggestion that the ratio of s to d increases as we 
approach the surface of the YBCO is due to Bahcall \cite{bahcall}. 
We use it to construct a simple model.  

We segment the sample into twins, which, 
if inclusions are neglected, may be
thought of as forming approximately a one-dimensional array. (See Fig.1 ).      
It is known from corner junction experiments \cite{vh}
that coupling across twin boundaries {\it in the bulk} maintains
one sign for the $d$-component throughout.  
The rotation of the $a$ and $b$ axes across the boundary
implies that the twins alternate from $d+s$ to $d-s$
relative to axes fixed in space.
The alternation will not necessarily hold {\it at the surface} if the 
gap is more strongly $s$-like.  
The coupling across the twins at the surface will
have the effect of aligning the $s$-component if its relative strength is
large enough.  In our model the intertwin coupling is
different in the bulk and the surface,
so we segment the sample into bulk and surface parts.
The resulting picture is a one-dimensional array of 
Josephson junctions at the surface, as shown in Fig. 1.  Each is 
characterized by the phase of its $s$-component, denoted by $\phi_i$ for the 
$i$th twin.
They are in an 'external field' exerted by the 
coupling to the bulk
below.  This external field has an antiferromagnetic character
due to the alternating nature of the 
twins from $d+s$ to $d-s$, and the fact that
only the $s$-component will couple to the surface.  
Denote the phase of the $s$-wave component 
in the bulk by $\phi^b_i$. 
The 'junction' between the bulk and the surface
is a complicated intervening region.  The total energy of this region
will depend on the twist of the order parameter in it, \ie, on $\phi_i - 
\phi_i^b$, and must be periodic in this difference; the lowest Fourier 
component has the standard form $ -J \cos(\phi_i - 
\phi_i^b)$.  Thus, in the model, the coupling appears Josephson-like.

Fig. 1 suggests a number of interesting effects.  For example, the couplings 
around a vertex which meets 4 separate regions, 2 surface and 
2 bulk, may give rise to phase 
frustration.  This permits the possibility of spontaneous flux running along 
the twin boundary.  If this occurs, it would tend to wash out the phase 
alternation on the surface \cite{andym} and to produce an overall Josephson 
coupling to the Pb.  We 
shall assume that the appropriate Josephson penetration depth is too short to 
permit this, but it is actually difficult to estimate.  Secondly, several 
papers \cite{belzig} have shown that, under certain circumstances, 
{\it current} may be expected along the boundary.  This is unlikely to affect
the c-axis tunneling results.  

In the absence of an applied field, the 
energy for this model of the YBCO surface is:
\begin{equation}
F = - \sum_{\langle ij \rangle} K_{ij} \cos(\phi_i-\phi_j)
    - \sum_i J_i \cos(\phi_i-\phi^b_i),
\end{equation}
where $\cos(\phi_i^b) = (-1)^i$.  The total 
Josephson current to the Pb electrode is 
given by
\begin{equation}
{\cal I} = \sum_i I_{ci} \sin(\phi_{Pb}-\phi_{i}), 
\end{equation}
where the Pb phase is taken to be uniform - the twin width is short 
compared to the Josephson penetration depth.
We restrict our considerations to low temperature, as experiments are 
only possible below the $T_c$ of Pb.

The ordered model is defined by the relations
$K_{ij} = K$, $J_i = J$, and $I_{ci} = I$.  The ratio of surface to bulk 
coupling is given by $K/J$.  The mean field solution in the regime $J<4K$ is
$\cos(\phi_i) = (-1)^i J/4K$, with a maximum Josephson current 
\begin{equation}
{\cal I}_{max} = N_T I_c \sqrt{1 -\frac{J^2}{16K^2}}.
\end{equation}
If $J>4K$, then $\cos(\phi_i) = (-1)^i $, and ${\cal I} = 0$. 
The supercurrent per unit area for a multitwin sample
is therefore reduced from the value it would have in an untwinned sample,
as is observed.

If a field H is applied parallel to the twin boundaries, the maximum 
supercurrent is found by maximizing the expression
\begin{equation}
{\cal I} = I_c \int \sin\left[\phi_{Pb}-\phi(x)+ \frac{Ht}{\Phi_0}x \right] dx 
\label{eq:ih}
\end{equation}
with respect to $\phi_{Pb}$.  Here $\Phi_0$ is the flux quantum and t
is the effective electrical junction width, including the penetration depths of 
the Pb and the YBCO.  For a single twin this of course gives 
the usual Fraunhofer pattern.  For two twins, there is a range
of possible behaviors.  For large $J/K$, there is a $180^{\circ}$ phase change 
and a symmetric field dependence:  ${\cal I}_{max}(H) = {\cal I}_{max}(-H)$ and
${\cal I}(H=0)=0$.  For intermediate $J/K$, we have the possibility of some 
asymmetry: ${\cal I}_{max}(H) \neq {\cal I}_{max}(-H)$ and some current even at 
zero field ${\cal I}(H=0) \neq 0$.  This last is due to breaking of 
time-reversal symmetry - the phase change across the boundary 
is less than $180^{\circ}$.  These two possibilities are illustrated for twins 
of equal areas in Fig. 2.
The experimental situation is likely to be close to the symmetric case.  The 
areas of twins are large and one may expect the coupling to the underlying bulk 
to dominate.  Experimentally, some asymmetry is indeed seen.  This may also be 
attributed partly to unequal twin areas or self-field effects 
\cite{agterberg} as well as time-reversal symmetry breaking found here.

A distinguishing feature of the present model is that, although time-reversal 
symmetry is broken by the competition between surface and bulk couplings, this 
does not necessarily show up in field-dependence experiments on multitwin samples.
The overall alternation of the two types of domains will tend to wash out the 
asymmetric part of the signal.  We illustrate this by a numerical solution of 
Eq.\ \ref{eq:ih} for a sample with 500 twins.  In Fig. 3(a) we show 
the field dependence of I$_c$ for an ordered sample with $J=2K$.  The pattern is 
the simple Fraunhofer one.  In Fig. 3(b), the same size sample is used, but the 
twin area varies about the same mean.  A gaussian distribution of fractional 
width 0.6 was chosen.  The coupling to the bulk for each twin varies 
proportionately.  The resulting field dependence shows no asymmetry at low 
fields.  At high fields smaller length scales are probed.  Because they are 
not subject to quite such complete averaging, asymmetry appears.  This is an 
experimental prediction of the model.

Disorder in general will reduce the Josephson current.
Some disorder is expected in every 
multitwin sample as the areas of the twins vary to a certain extent, and the 
coupling to the bulk should increase as the area increases.  
This is illustrated in Fig. 4 for a sample with 100 twins and a distribution of 
couplings to the bulk.  For a very disordered sample, no supercurrent can 
flow.  Detailed comparison of theory and experiment in this regard is
unfortunately difficult.  The points in Fig. 4 correspond to different samples.
In practice, samples with different twin densities have
contact resistances which vary over an order of magnitude, showing that
surface quality is highly depemdent on preparation methods.  

Our results may be summarized by a magnetic analogy.  Think of the phases of the 
$s$-components at the surface as permanent magnetic moments confined to a plane.  
The surface coupling is ferromagnetic, while the coupling to the bulk gives 
rise to an antiferromagnetic external field.  
The maximum supercurrent at zero field represents 
the resultant moment.  The maximum supercurrent at finite applied magnetic 
field represents the moment at a nonzero wavevector.  
When the surface coupling dominates the coupling of 
surface and bulk, we have ferrimagnetism and a nonzero resultant moment.  
When the randomness in the exernal field becomes large, 
the individual moments become random in direction and there is no resultant.  
 
The overall picture which results is that of nearly pure $d$-wave in the 
bulk and mixing of $s$-wave and $d$-wave near the surface, with 
$s$-wave perhaps predominating.  This picture may well be 
specific to YBCO.  In BSCCO-type high-T$_c$ compounds, for example,
it seems likely that the surface 
(as seen in photoemission) is representative of 
the bulk, and both are $d$-wave, at least in the optimally doped case.
It is worth noting that the any disordered model which includes mixing will
exhibit some sort of time-reversal symmetry breaking.  No gauge transformation 
can restore all of the disordered phases to zero.  Whether this breaking is 
visible in any particular experiment is a detailed question.

We are grateful to A.S. Katz,
D. Agterberg, R. Hlubina, R.B. Laughlin
and A.J. Leggett for helpful discussions. 
This work is supported by the NSF under the
Materials Research Science and Engineering Center Program,
Grant No. DMR-96-32527 and by DMR-9704972.

\begin{figure}
\caption[]{(a): schematic representation of the surface
of a twinned sample, viewed along the c-axis.
(b): the twin structure can be viewed as a one-dimensional
alternation of different domains if the inclusions are ignored.}
\end{figure}

\begin{figure}
\caption[]{The effects of an applied field on a junction with 
two twins of equal area for parameter values
(a): $J>4K$, and (b): J=2K.
A large coupling to the bulk 
gives a symmetric field dependence, while a weaker bulk
coupling results in asymmetry.}
\end{figure}

\begin{figure}
\caption[]{(a): an ordered junction with J/K=2 and 500 
twins of equal area.  We obtain a symmetric field dependence, regardless 
of the time reversal breaking state which 
exists in the surface layer.  (b): when a gaussian distribution
of twin sizes and couplings J/K is used,
asymmetry appears at higher fields.}
\end{figure}

\begin{figure}
\caption[]{$I_c (H=0)$ for a junction of 100 twins 
with a gaussian distribution
of bulk couplings J$_i$ as a function of distribution width.  
As the disorder increases, $I_c(H=0)$ is diminished.}
\end{figure}  
 

\begin{references}
\bibitem{note} This would most likely
not be the case if the $s$-wave part
is purely secondary, and implies that $s$-wave and 
$d$-wave must be somewhat competitive energetically even in the absence of 
orthorhombicity.
\end{references}
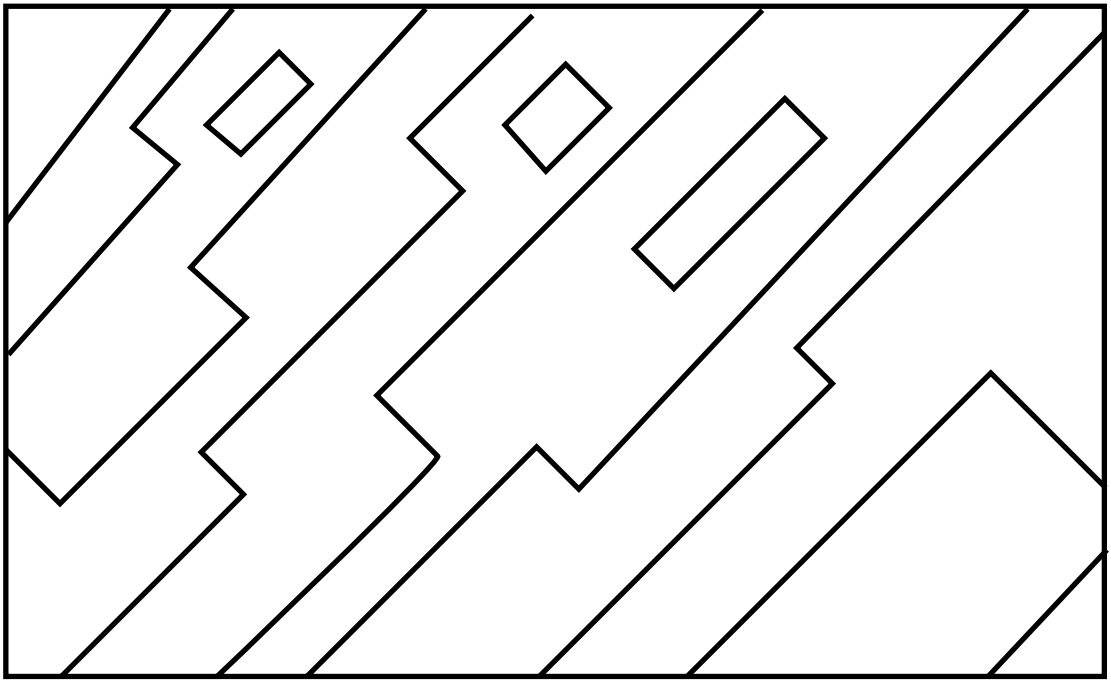
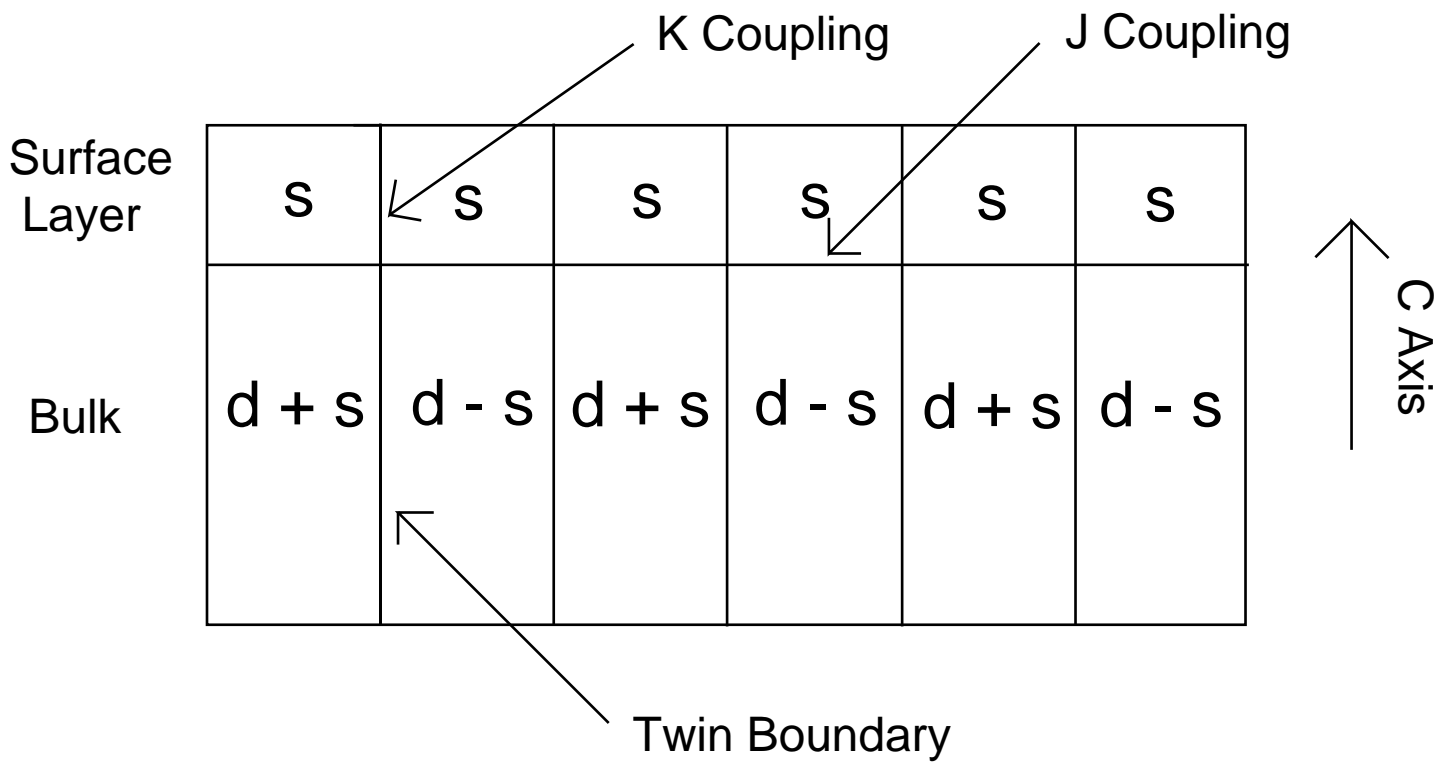
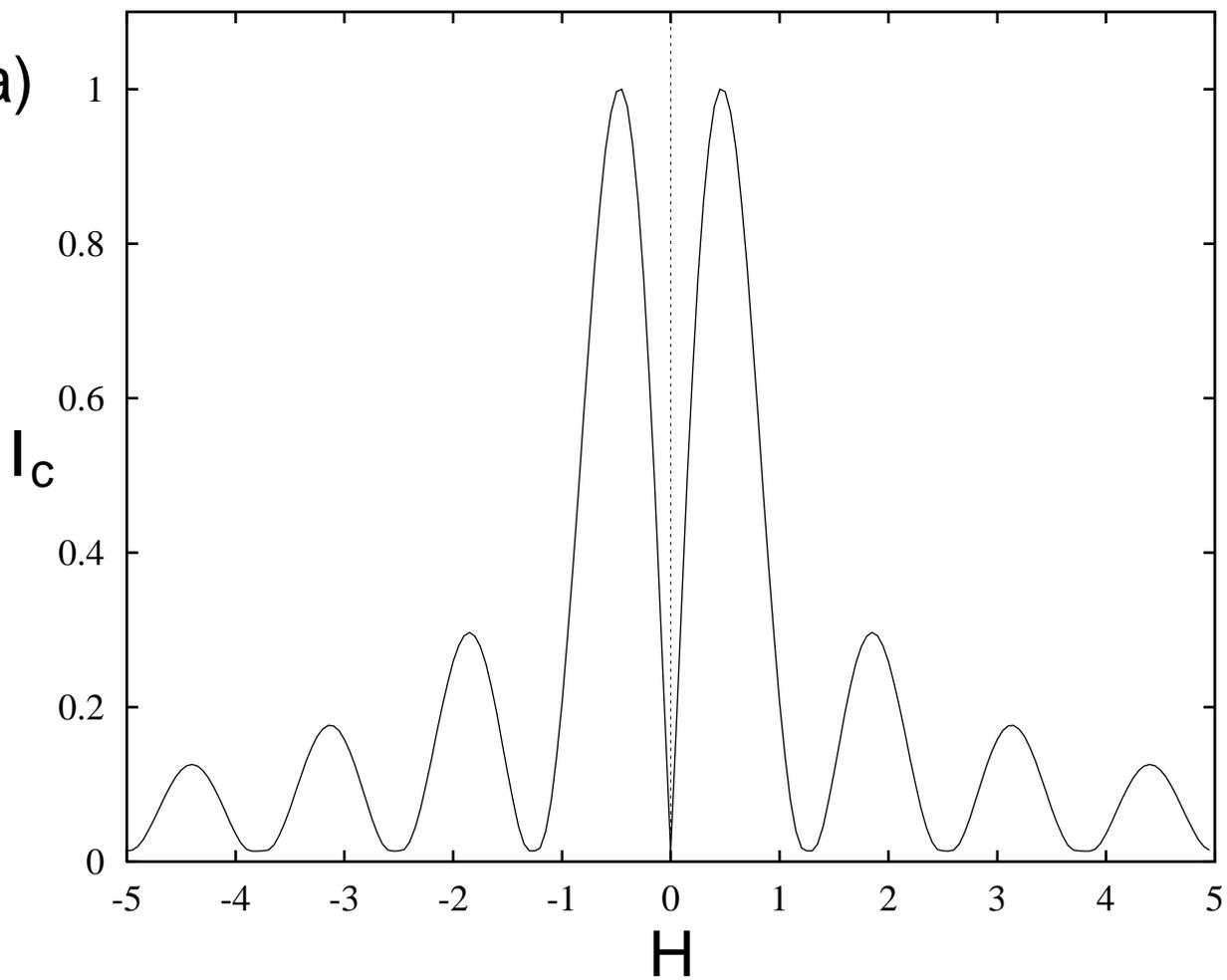
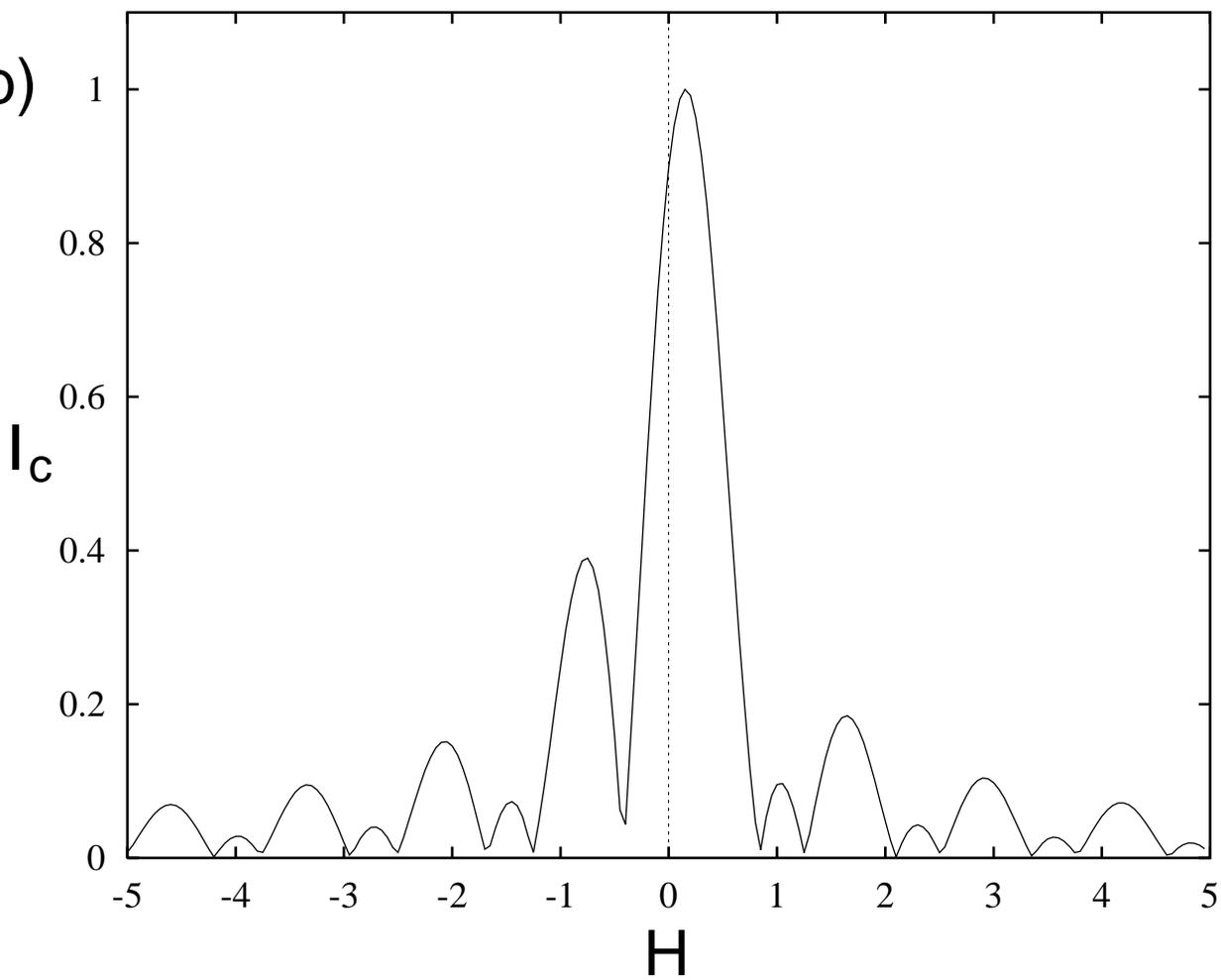
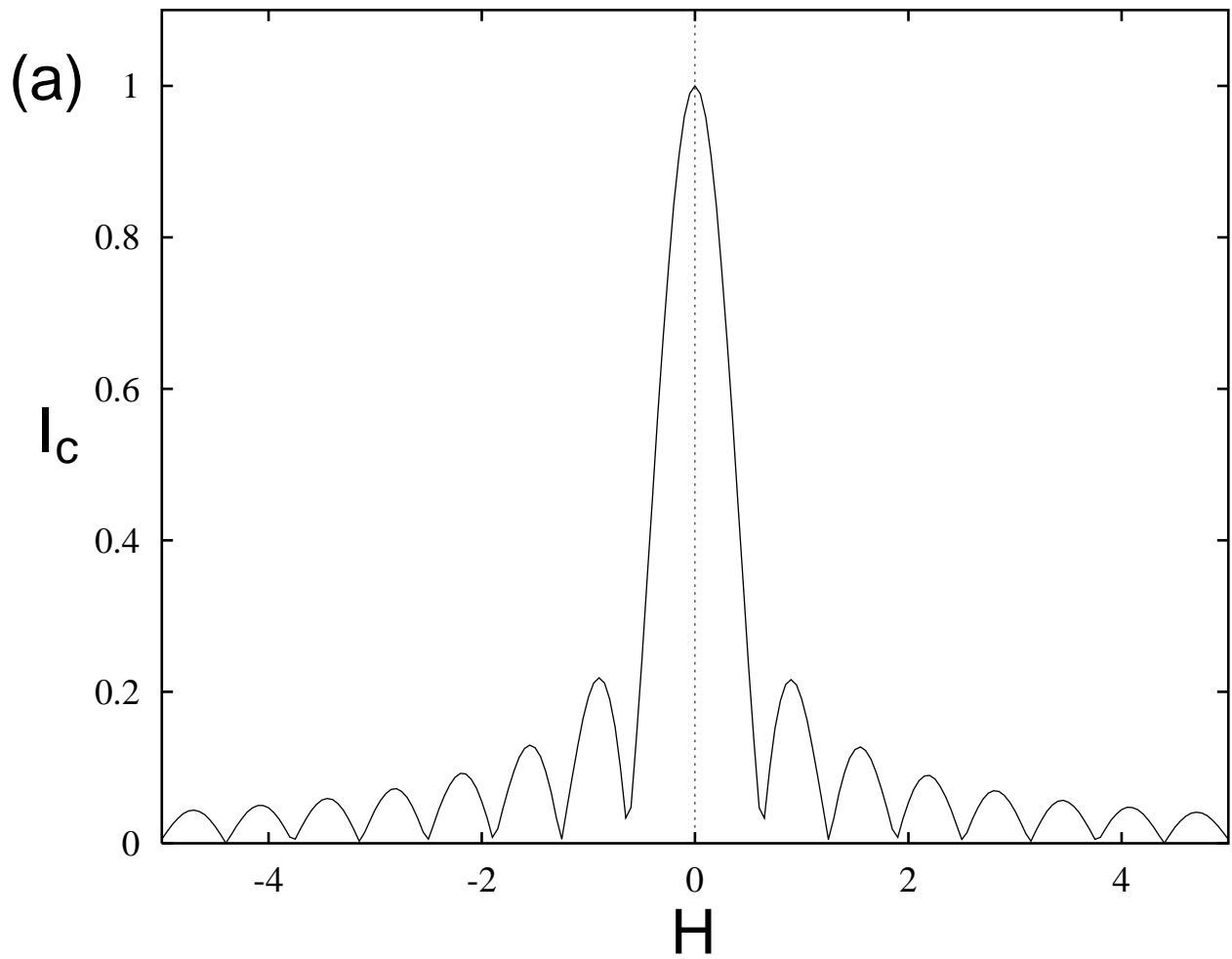
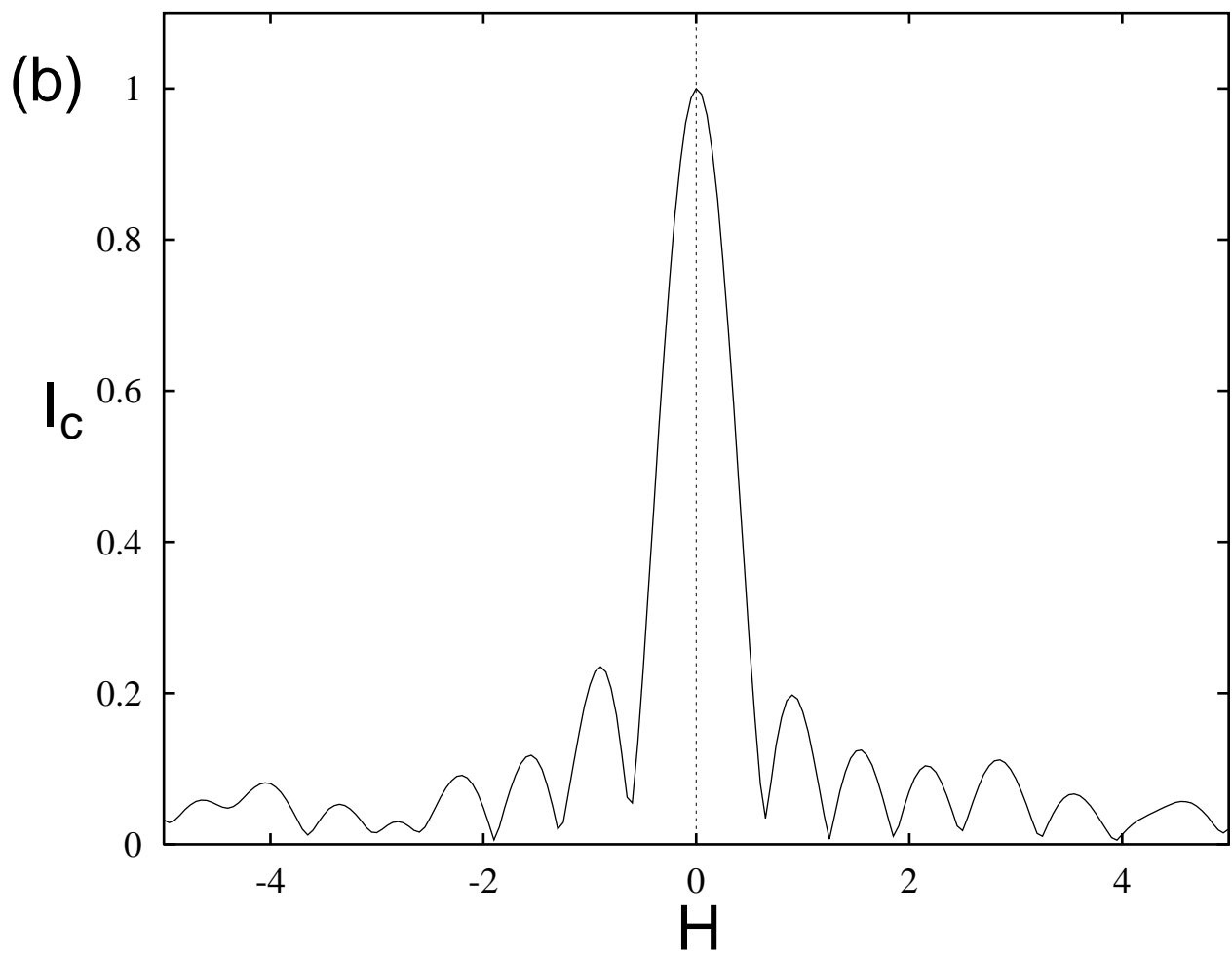
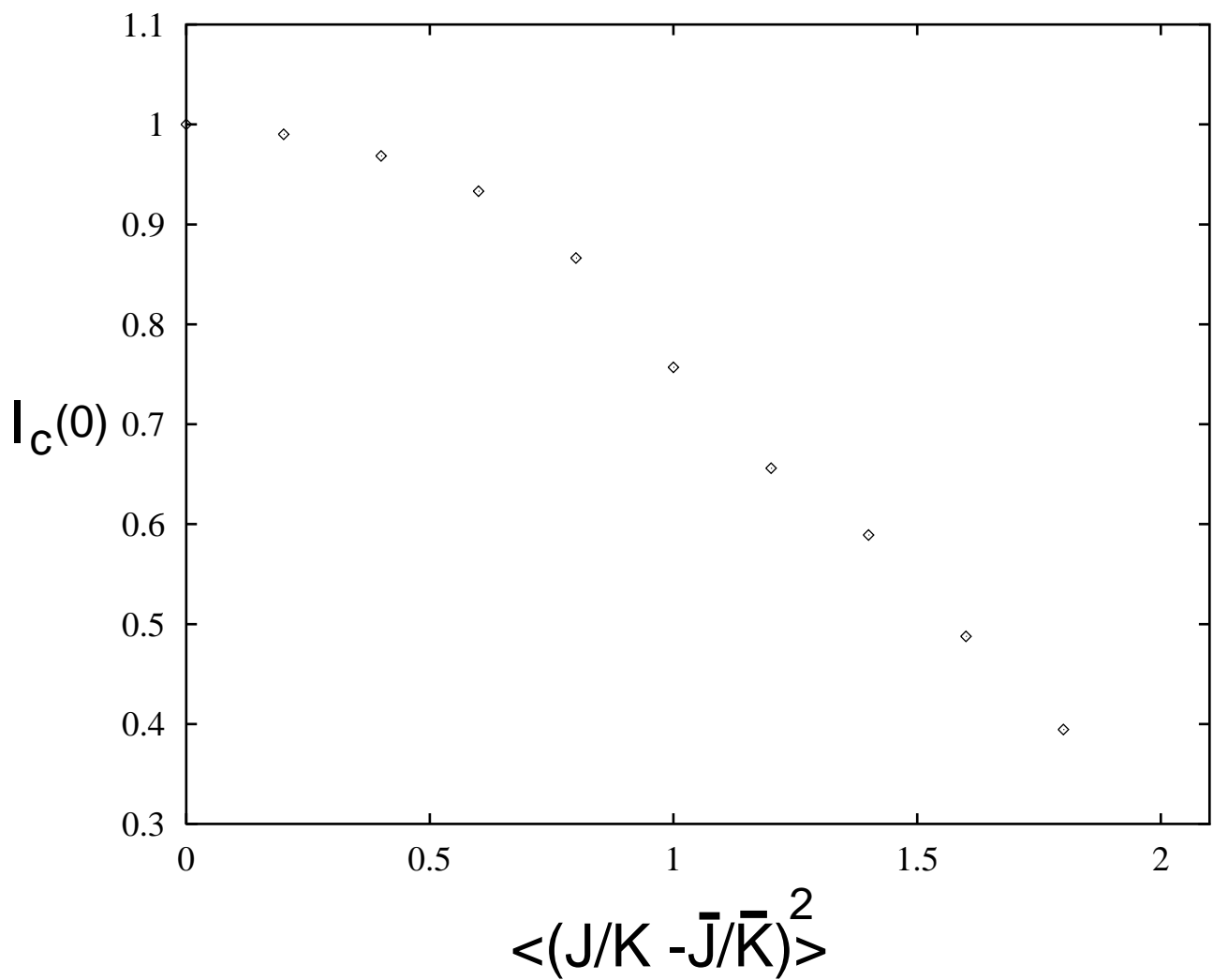
\end{document}